\documentclass[prd,twocolumn,superscriptaddress,amsfonts,amssymb,amsmath,showpacs,showkeys]{revtex4-1}

\usepackage{graphics}
\usepackage{eurosym}
\usepackage{amsfonts}
\usepackage{amssymb}
\usepackage{amsmath}
\usepackage{graphicx}
\usepackage[font={footnotesize,it}]{caption}
\usepackage{hyperref}
\usepackage{float}

\begin{document}

\title{Scalar Particles¡¯ Tunneling Radiation in the Demianski-Newman Spacetime
with Influences of Quantum Gravity}

\author{ Zhonghua Li}
\email{sclzh888@163.com}
\affiliation{Institute of Theoretical Physics, China West Normal University,Nanchong, Sichuan 637002, China}

\date{\today }

\begin{abstract}
 In this paper,using Hamilton-Jacobi ansatz, we investigate scalar particles¡¯
    tunneling radiation in the Demianski-Newman spacetime . We get the effective temperature with influences of quantum gravity, and compare this temperature with the original temperature of the Demianski-Newman
     black hole. We find that it's similar to the case of fermions, for scalar particles the influence of quantum gravity will also slow down the increase of Hawking temperatures,which naturally leads to remnants left in the evaporation.

\end{abstract}

\maketitle

\section{Introduction}

Hawking proposed that black holes exist radiation. The study of the black hole radiation is one of the important directions of black hole physics. WKB approximation is usually used to calculate the tunneling rate of emitted particles \cite{APGS2}
\begin{eqnarray}
 \Gamma &\propto& exp\left[-Im\oint p_{r} dr\right] \nonumber \\
 &=& exp \left[-Im (\int p^{out}_{r}dr -\int p^{in}_{r}dr)\right]
\end{eqnarray}
The closed path above goes across the horizon and comes back.
Related discussion about the tunneling rate of emitted particles in detail in \cite{ETA1,ETA2,ETA3}.

The Null Geodesic Method can be used to calculate the imaginary part of emitted particle¡¯s action. It was used in the work of Parikh and Wilczek \cite{PW}. In this method,the Painleve coordinate transformation should be performed. The main characters of these coordinates are their stationary and nonsingular around the horizon. So one can obtain the imaginary part by canonical momenta and Hamilton canonical equations. The other method which can be used to calculate the imaginary part of emitted particles's action is the Hamilton-Jacobi Ansatz \cite{ANVZ} which was first proposed in \cite{SP,SPS}.In this method, the action of the system accords with the Hamilton-Jacobi equation.Taking into account the property of the spacetime, one can make a separation of variables on the action $I=-\omega t+W(r)+\Phi(\phi,z)$.Then inserting the results into the Hamilton-Jacobi equation , one can acquire the imaginary part. Many researchers extend this work to the tunnelling radiation of fermions. The standard Hawking temperatures of the spherically symmetric and charged black holes were discussed in \cite{KM2}. Other work about fermions' tunnelling radiation is referred to \cite{KM3,AS,AS1,LR,CJZ,QQJ,HCNVZ,HCNVZ1,BRM}.  Ref.\cite{BAM1} discussed a correction to the tunneling probability by taking into account the back reaction effect. In Ref.\cite{BMS} features regarding the variation of the Hawking temperature,entropy and tunneling rate were revealed, including corrections due to non-commutativity and back reaction. Ref.\cite{BAM2,BAM3} discussed the quantum tunneling beyond semiclassical approximation and  trace anomaly by Hamilton-Jacobi method. Ref.\cite{MAS} discussed the Hawking radiation due to photon and gravitino tunneling. Topics related to the entropy-area spectrum of a black hole were discussed in Ref.\cite{BAM5,BMAV1,BRM1}. Other important advances are referred to \cite{BAM4,BMAV2,BKAM,KBAM}.

Taking into account the theory of quantum gravity, there exists  a minimal observable length \cite{PKTC,PKTC1,PKTC2,PKTC3,PKTC4}. Then there is  the generalized uncertainty principle (GUP)

\begin{eqnarray}
\Delta x \Delta p \geq \frac{\hbar}{2}\left[1+ \beta \Delta p^2\right],
\label{eq1.1}
\end{eqnarray}

\noindent here $\beta = \beta_0 (l^2_p /\hbar^2)$, $\beta_0 $ is a dimensionless parameter and $l_p$ is the Planck length. The derivation of the GUP is based on the modified fundamental commutation relations. Kempf et al. first modified commutation relations \cite{KMM} and got $\left[x_i,p_j\right]= i \hbar \delta_{ij}\left[1+ \beta p^2\right]$, where $x_i$ and $p_i$ are operators of position and momentum defined by

\begin{eqnarray}
x_i = x_{0i}, p_i = p_{0i} (1 + \beta p^2),
\label{eq1.2}
\end{eqnarray}

\noindent and here $x_{0i}$ and $p_{0i}$ satisfy the basic quantum mechanics commutation relations $\left[x_{0i},p_{0j}\right]= i \hbar \delta_{ij}$.

 With consideration of the GUP, the cosmological constant problem was discussed, and the finiteness of the constant was derived in \cite{CMOT}. Based on the new form of GUP, the Unruh effect has been discussed in \cite{BM06}. The quantum dynamics of the Friedmann-Robertson-Walker universe was derived in \cite{BM1}. Some predictions on post inflation preheating in cosmology were made in \cite{CDAV}. Based on the generalization, the thermodynamics of the black holes were discussed again in \cite{KSY,XW,BJM} .

When effects of quantum gravity are taken into account,we investigated fermions¡¯ tunnelling from the charged and rotating
black strings\cite{CL}. We found that quantum gravity corrections slow down the increases of the temperatures, which naturally leads to remnants left in the evaporation. In \cite{LZ}, we discussed the tunnelling of fermions when effects of quantum gravity are taken into account. We investigated two cases, black string and Kerr AdS black hole. We found that For black string, the uncharged and un-rotating case,the correction of Hawking temperature is only affected by the mass of emitted fermions and the quantum gravitational corrections slow down the increases of the temperature, which naturally leads to remnants left in the evaporation,too. For another case, the Kerr AdS black hole, we found that the quantum gravitational corrections are not only determined by the mass of the emitted fermions but also affected by the rotating properties of the AdS black hole. So with consideration of the quantum gravity corrections, an offset around the standard temperature
always exists.

In this paper, we do not discuss fermions but focus on scalar particles¡¯ tunneling radiation in the Demianski-Newman spacetime
with influences of quantum gravity. The Demianski-Newman black hole is similar to Kerr-Newman black hole. It is an extension of Kerr-Newman black hole \cite{GPZ}. Its expression of metric is more complex than Kerr-Newman black hole, so the calculation of rate of tunneling radiation is not easy relatively. Using Hamilton-Jacobi ansatz, we investigate scalar particles¡¯ tunneling radiation  in the Demianski-Newman spacetime. With consideration of quantum gravity, we get the effective temperature,
    and compare this temperature with the original temperature of the Demianski-Newman
     black hole. We find that it's similar to the case of fermions, for scalar particles the quantum gravity effects also slow down the increase of Hawking temperatures.

The rest is organized as follows. In the next section, based on the modified commutation relations of  operators of position and momentum defined in \cite{KMM}, we will derive the Generalized Klein-Gordon  equation in curved spacetime. In Section 3, we investigate Scalar particles¡¯ tunneling radiation in the
  Demianski-Newman black hole. Section 4 is devoted to our conclusion.

\section{Generalized Klein-Gordon  equation in curved spacetime}

In this section, to accunt for the effects from quantum gravity ,we first write the minimal length which is generalized
   uncertainty principle as
   \begin{eqnarray}
\Delta x \Delta p \geq \frac{\hbar}{2}\left[1+ \beta \Delta p^2\right],
\label{eq1.1}
\end{eqnarray}

\noindent then we investigate the Klein - Gordon equation in curved space-time
 with the effects of GUP . We adopt the position
 representation,and take the effect of quantum gravity into
 consideration ,the momentum operators are written as
 \begin{eqnarray}
x_i = x_{0i},p_{0i}=i\hbar\partial_{i}\hspace{5mm} p_i = p_{0i} (1 + \beta p^2).
\label{eq1.2}
\end{eqnarray}
\noindent Because $ \beta $ is a small quantity, so we neglect the high orders of $ \beta $,and get the
 square of  momentum operators

 \begin{eqnarray}
p^2 &=& p_i p^i = -\hbar^2 \left[ {1 - \beta \hbar^2 \left( {\partial _j \partial ^j} \right)}
\right]\partial _i \cdot \left[ {1 - \beta \hbar^2 \left( {\partial ^j\partial _j } \right)}
\right]\partial ^i\nonumber \\
&\simeq & - \hbar ^2\left[ {\partial _i \partial ^i - 2\beta \hbar ^2
\left( {\partial ^j\partial _j } \right)\left( {\partial
^i\partial _i } \right)} \right].
\label{eq2.1}
\end{eqnarray}
To take into account effects of quantum gravity, generalize the frequency as
\begin{eqnarray}
\tilde \omega = E( 1 - \beta E^2),
\label{eq2.2}
\end{eqnarray}

\noindent where the energy operator takes the form of $ E=i\hbar\partial_{0} $ . According to  the energy-mass shell condition: $ p^{2}+m^{2}=E^{2} $ ,  the modified expression of energy can be written as
\begin{eqnarray}
\tilde E = E[ 1 - \beta (p^2 + m^2)].
\label{eq2.3}
\end{eqnarray}

\noindent So we get the expression of Klein-Gordon equation in the curved space-time
\begin{eqnarray}
g^{\mu\nu}p_{\mu}p_{\nu}\Psi =-m^{2}\Psi.
\label{eq2.4}
\end{eqnarray}
\noindent To take into account the effects of quantum gravity,the above equation take the form as
\begin{eqnarray}
-(i\hbar)^{2}\partial^{t}\partial_{t}\Psi=[(i\hbar)^{2}\partial^{i}\partial_{i}+m^{2}]\Psi.
\label{eq2.5}
\end{eqnarray}
\noindent According to  the above expression and keeping the foremost order of $ \beta $,  the generalized Klein-Gordon equation is rewritten as
\begin{eqnarray}
-(i\hbar)^{2}\partial^{t}\partial_{t}\Psi\nonumber \\
=[(i\hbar)^{2}\partial^{i}\partial_{i}+m^{2}][1-2\beta((i\hbar)^{2}\partial^{i}\partial_{i}+m^{2})]\Psi
\label{eq2.6}
\end{eqnarray}

\section{Scalar particles¡¯ tunneling radiation in the
  Demianski-Newman black hole}

In this section ,we research the tunnelling radiation of Scalar
 particles in the Demianski-Newman black hole by useing the above
  generalised Klein-Gordon equation. The metric is given by \cite{GPZ}
\begin{eqnarray}
  ds^2=-\frac{\Delta-a^{2}sin^{2}\theta}{\Sigma} dt^2+\frac{\Sigma}{\Delta}dr^2+\Sigma d\Theta^{2}-\nonumber \\
  \frac{2(aBsin^{2}\theta-A\Delta)}{\Sigma}dtd\varphi+\frac{B^{2}sin^{2}\theta-A\Delta}{\Sigma}d\varphi^{2}
\label{eq2.7}
\end{eqnarray}
Where \\
\begin{eqnarray}
\Delta=r^{2}-2Mr+a^{2}-l^{2}\hspace{5mm}\nonumber \\
B=r^{2}+a^{2}+l^{2}\nonumber \\
A=asin^{2}\Theta+2lcos\theta \hspace{5mm}\nonumber \\
 \Sigma=r^{2}+a^{2}cos^{2}\theta+l^{2}-2alcos\theta.
\label{eq2.8}
\end{eqnarray}
\noindent Here $a$ reprents angular momentum per unit mass and $M$ reprents the black hole¡¯s mass. Parameter $l$ in  Demianski-Newman metric is  an extended parameter relative to Kerr-Newman black hole(when $l=0$, the Demianski-Newman black hole reduces to Kerr-Newman black hole), it is related to rotation effect and it can generate potential field. Ref.\cite{GPZ} inferred that it may be some charge concerning spin.  Black hole have inner and outer horizion ,we only consider the outer horizion:  $ r_{h}^{+}=M+\sqrt{M^{2}-a^{2}+l^{2}} $. The entropy of black hole is defined as :  $ S=\pi(r_{h^{+}}^{2}+a^{2}+l^{2}) $. Because the black hole rotates, it is not easy to research the particle¡¯s tunneling directly, so we take the above metric dragging transformation as follow
\begin{eqnarray}
\phi=\varphi-\Omega t
\label{eq2.9}
\end{eqnarray}
Where
\begin{eqnarray}
\Omega=\frac{aBsin^{2}\theta-A\Delta}{B^{2}sin^{2}\theta-A^{2}\Delta}
\label{eq3.0}
\end{eqnarray}
Then we get :
\begin{eqnarray}
ds^2=-\frac{\Delta\Sigma}{(r^{2}+a^{2}+l^{2})^{2}-\frac{\Delta A}{sin^{2}\theta}}dt^2+\frac{\Sigma}{\Delta}dr^2+\Sigma d\theta^{2}-\nonumber \nonumber \\
+\frac{(r^{2}+a^{2}+l^{2})^{2}-(asin^{2}\theta+2lcos\theta )^{2}\Delta}{\Sigma}d\phi^{2}.\nonumber \\
\label{eq3.1}
\end{eqnarray}
\noindent To solve the generalised Klein-Gordon equation, we sppose the wave function of the scalar particle takes the form as
\begin{eqnarray}
\Psi=exp[\frac{i}{\hbar}I(t,r,\theta,\phi)]
\label{eq3.2}
\end{eqnarray}
\noindent In the wave function, $I$ represents the action of the scalar particle. Inserting the wave function into the generalised  Klein-Gordon equation  and the metric in Demianski-Newman black hole,  considering the WKB approximation, keeping the leading order of $ \beta $ and $ \hbar $ and neglecting the higher  orders of them, we get
\begin{eqnarray}
\frac{1}{F}(\partial_{r}I)^{2}=[(G(\partial_{r}I)^{2}+\frac{1}{K^{2}}(\partial_{\theta}I)^{2}+\frac{1}{H^{2}}(\partial_{\phi}I)^{2})+m^{2}]\nonumber \nonumber \\
(1-2\beta[(G(\partial_{r}I)^{2}+\frac{1}{K^{2}}(\partial_{\theta}I)^{2}+\frac{1}{H^{2}}(\partial_{\phi}I)^{2})+m^{2}]),\nonumber \\
\noindent H^{2}=\frac{(r^{2}+a^{2}+l^{2})^{2}-(asin^{2}\theta+2lcos\theta )^{2}\Delta}{\Sigma},\hspace{5mm}K^{2}=\Sigma.\nonumber \\
\label{eq3.3}
\end{eqnarray}

\noindent To get the solution of the above equation directly is not easy. In order to take separation of variables easily, we consider the form of the action of the scalar particle as follow
\begin{eqnarray}
I=-(\omega-j\Omega) t+ W\left(r\right) + j(\theta,\phi)
\label{eq3.4}
\end{eqnarray}
\noindent
where $ \omega $ represents the energy of the emitted particle.We insert the above expression of action $I$  into the generalised  Klein-Gordon equation  by using separation of variables, then we get
\begin{eqnarray}
 C_4\left({\partial _r W} \right)^4 +C_2\left( {\partial _r W} \right)^2 + C_0 = 0
\label{eq3.5}
\end{eqnarray}
where
\begin{eqnarray}
C_4 & = &-2 \beta G^{2}\nonumber \\
C_2 & = & G(1-4\beta m^{2})\nonumber \\
C_0 & = & m^{2}-2\beta m^{4}-\frac{(\omega-j\Omega)^{2}}{F}.
\label{eq3.6}
\end{eqnarray}
\noindent Then we can get action of radial direction as follow
\begin{eqnarray}
W_{\pm} = \pm\int \frac{dr}{\sqrt{GF}} \sqrt{\left( \omega -j\Omega\right)^2-m^{2}F}\nonumber \\
\left(1+\beta m^{2} + \beta\frac{(\omega - j\Omega)^2}{F}\right)\nonumber\\
= \pm i\pi(\omega - j\Omega_{h^{+}})\frac{r_{h^{+}}^2+a^2+l^{2}}{r_{h^{+}}-r_{h^{-}} }\left(1+ \beta \xi \right)+(real\hspace{3mm} part).\nonumber \\
\label{eq3.7}
\end{eqnarray}
\noindent In above equation $+/-$ represents outward/ interior solution,respectively,\nonumber \\
where
\begin{eqnarray}
\Omega_{h^{+}}=\frac{a}{r_{h^{+}}^2+a^2+l^{2}}\hspace{5mm}r_{h^{+}}=M+\sqrt{M^{2}-a^{2}+l^{2}}\nonumber \\
F=\frac{\Delta\Sigma}{(r^{2}+a^{2}+l^{2})^{2}-\frac{\Delta A^{2}}{sin^{2}\theta}}\hspace{5mm}G=\frac{\Delta}{\Sigma}\nonumber \\
\label{eq3.8}
\end{eqnarray}
\noindent The expression of $ \xi $ is pretty complicated, so we do not write it here .Then ,with WKB appoximation, the tunnelling rate of the scalar particle can be writted as
\begin{eqnarray}
\Gamma=\frac{P_{out}}{P_{in}}=\frac{exp(-2I_{m}I_{+})}{exp(-2I_{m}I_{-})}\nonumber \\
=exp[-2(I_{m}I_{+}-I_{m}I_{-})]\nonumber \\
=exp[4\pi(\omega - j\Omega_{h^{+}})\frac{r_{h^{+}}^2+a^2+l^{2}}{r_{h^{+}}-r_{h^{-}} }\left(1+ \beta \xi \right)].
\label{eq3.19}
\end{eqnarray}
\noindent According to the expression of the Bolzmannn factor ,we can get the corrected Demianski-Newman temperature as :
\begin{eqnarray}
T=\frac{r_{h^{+}}-r_{h^{-}}}{4\pi(r_{h^{+}}^2+a^2+l^{2})(1+\beta\xi)}\nonumber \\
=\frac{T_{0}}{1+\beta\xi}=T_{0}(1-\beta\xi),
\label{eq3.20}
\end{eqnarray}
\noindent where the original temperature of the Demianski-Newman black hole is
 \begin{eqnarray}
 T_{0}=\frac{r_{h^{+}}-r_{h^{-}}}{4\pi r_{h^{+}}^2+a^2+l^{2}}
 \label{eq3.21}
\end{eqnarray}

\section{Discussion and conclusions}
In this paper ,we discuss Scalar particles¡¯ tunneling radiation in the Demianski-Newman spacetime with the effects of quantum gravity. According to the effective Demianski-Newman temperature,apparently ,we can conclude that the corrected Demianski-Newman temperature is lower than the original temperature of the Demianski-Newman black hole.Because $ \beta $ is a small quantity, it's similar to the case of fermions, for scalar particles the influence of quantum gravity will also slow down the evaporation of the Demianski-Newman black hole, so when the black hole is at the balance point ,there is remnants in the black hole.

\begin{acknowledgments}

This work is supported by the Fundamental Research Funds of China West Normal University (13C009).We thank L. M. Zhang for her participation in the early stage of this work.

\end{acknowledgments}

\end{document}